\pdfoutput=1

\documentclass[11pt]{article}

\usepackage[]{EMNLP2023} 
\usepackage{graphicx}
\usepackage{times}
\usepackage{latexsym}
\usepackage[english]{babel}
\usepackage{amsmath}
\usepackage{hyperref}

\usepackage{amsthm}
\usepackage{algorithm}
\usepackage{algpseudocode}
\usepackage{graphicx}
\usepackage{subcaption}
\usepackage{amssymb}

\makeatletter
\title{The Silent Curriculum: How Does LLM Monoculture Shape Educational Content and Its Accessibility?}
\author{Aman Priyanshu\thanks{* Equal contribution} \\ School of Computer Science \\ Carnegie Mellon University \\ (apriyans@andrew.cmu.edu) \And
\textbf{Supriti Vijay}\footnotemark[1] \\ Adobe, India \\ (supriti.vijay@gmail.com)}
\makeatother
\begin{document}

\maketitle

\begin{abstract}
As Large Language Models (LLMs) ascend in popularity, offering information with unprecedented convenience compared to traditional search engines, we delve into the intriguing possibility that a new, singular perspective is being propagated. We call this the "Silent Curriculum," where our focus shifts towards a particularly impressionable demographic: children, who are drawn to the ease and immediacy of acquiring knowledge through these digital oracles. In this exploration, we delve into the sociocultural ramifications of LLMs, which, through their nuanced responses, may be subtly etching their own stereotypes, an algorithmic or AI monoculture. We hypothesize that the convergence of pre-training data, fine-tuning datasets, and analogous guardrails across models may have birthed a distinct cultural lens. We unpack this concept through a short experiment navigating children's storytelling, occupational-ethnic biases, and self-diagnosed annotations, to find that there exists strong cosine similarity (0.87) of biases across these models, suggesting a similar perspective of ethnic stereotypes in occupations. This paper invites a reimagining of LLMs' societal role, especially as the new information gatekeepers, advocating for a paradigm shift towards diversity-rich landscapes over unintended monocultures.
\end{abstract}

\section*{Introduction: What is LLM Monoculture?}

Imagine a world where the vast expanse of human knowledge and cultural diversity is distilled through a singular digital prism, presenting a streamlined, yet fundamentally altered, vision of our world. This is precisely the reality unfolding with LLMs, where the convergence of pre-training data, fine-tuning datasets, and uniform de-biasing techniques coalesce into what we might perceive as a unified cultural lens, subtly crafting a new, yet undeniably distinct, digital narrative. This phenomenon, which we dub "LLM monoculture," emerges not from a deliberate act of cultural homogenization but from the intrinsic mechanics of LLMs \cite{Kleinberg_2021, NEURIPS2022_17a234c9}. Large AI organizations, in their quest to harness the internet's collective knowledge, inadvertently converge towards a uniform, distinct perspective. It's a perspective that, while not mirroring any single human culture, forms its unique narrative fabric.

However, the primary question this paper poses is not an exploration of what these perspectives entail but rather how they might harm society by becoming beacons of misinformation or arbiters of stereotypes. As LLMs ascend to become the primary source of knowledge, especially for the younger, digitally-native generation, they may inadvertently assume the role of educators. Children, drawn to the ease and immediacy of conversing with these all-knowing LLMs, may be introduced to what we term the "Silent Curriculum." This curriculum is not defined by what is explicitly taught but by what is implicitly conveyed through the models' responses. As these children query the LLMs, seeking answers and understanding, they may be unknowingly exposed to a filtered spectrum of cultural perspectives---one shaped by the underlying monoculture of the LLMs \cite{lalor2022benchmarking}. A monoculture inherently created by alignment, fine-tuning, and guardrailing techniques \cite{power1987monoculture}. This "Silent Curriculum" thus holds the capacity to become a pervasive force, subtly influencing the norms, biases, and stereotypes that the next generation may carry forward. In recognizing the potential of these LLMs to sculpt the future generations, we conduct a short exploratory study on occupational-ethnic biases.

\section*{An Empirical Gaze into the Silent Curriculum}

In our empirical exploration, we utilized two prominent LLMs, GPT-3.5 and LLaMA2-70B, as subjects to investigate the nuanced dynamics of cultural representation and occupational stereotypes. Initially, both models were prompted to draw inspiration from the WinoBias dataset to construct an Ethnicity and Top 20 Occupations corpus for a similar benchmark \cite{zhao2018gender, manela2021stereotype}. Remarkably, without external guidance, the models identified seven overlapping ethnic groups: White, Black, Asian, Hispanic, Native American, Middle Eastern, and Latin American, which formed the basis of our experimental framework (GPT3.5 generated 8, whereas LLaMA2-70B generated 7). Subsequently, we tasked the LLMs with generating short stories centered around a child's name and birthplace, leading to their eventual success in these given occupations, without specifying any ethnicity. This setup allowed us to examine the implicit ethnic and geographical biases present in the models' outputs. The final phase of our experiment involved extracting the names and locations from these stories to infer their ethnicities, as annotated by the LLMs themselves. Our findings revealed significant cosine similarities---0.86 for self-annotated ethnicity of the main character's name and 0.87 for the self-annotated ethnicity of the country---indicating a strong consistency in cultural representation across both models. We present detailed experimental setup and results in the Appendix~\ref{app:ed}. Notably, the models demonstrated a marked preference for characters belonging to Asian, White, Hispanic, and Black groups, with minimal representation of Latin American, Middle Eastern, and Native American identities.

\section*{Conclusion: Beyond Homogenization}

We posit that these results hint at underlying cultural nuances, sculpted not only by existing societal stereotypes but also by a potentially new set of cultural biases emerging from the LLM monoculture, offering a narrowed lens through which these digital entities view and narrate our world. We encourage the research community to dive deeper into these findings, seeking diverse perspectives to enrich our understanding of LLMs' cultural impact. This effort calls for a collective push from both academia and industry to challenge and broaden the AI monoculture, ensuring a multiplicity of voices shapes the responses guiding our future.


\bibliography{custom}
\bibliographystyle{acl_natbib}

\section*{Appendix}

\subsection*{Experimental Details}

We provide a comprehensive overview of the methodology and analytical processes employed in our exploration of Large Language Models (LLMs) and their potential to foster a monoculture, particularly focusing on their influence on educational content and accessibility through what we have termed the "Silent Curriculum." We observe this through the generation of Children's Short Stories. Our experiment utilized two advanced LLMs, GPT-3.5 developed by OpenAI and LLaMA2-70B developed by Meta, to investigate the nuances of cultural representation and occupational stereotypes. While we'd like to have extended it to other LLMs, we wished to discuss these preliminary findings through this Provocations Track.

\subsubsection*{Generating an Occupational-Ethnic Bias Benchmark}

The first phase of our experiment aimed at establishing a benchmark for occupational-ethnic biases present within LLMs. To accomplish this, we initially prompted GPT-3.5 and LLaMA2-70B to derive inspiration from the WinoBias dataset to create an occupational-ethnic bias dataset. This step was critical for guiding the models in creating a comprehensive Ethnicity and Top 20 Occupations corpus autonomously. The two models generated a list of eighth and seven ethnic groups respectively. Remarkably, seven of the groups generated by LLaMA2-70B was also generated by GPT3.5 models, there were: White, Black, Asian, Hispanic, Native American, Middle Eastern, and Latin American.

Following this, we asked both models to observe the WinoBias dataset's male \& female occupation lists and generate similar lists for each of the ethnic groups. We hypothesize that the research rooted initiative for this prompt, may have pushed these models to return responses without any guardrails or alignment acting up. For instance, among the occupations associated with the White ethnic group, roles such as "Corporate Executive," "Surgeon," and "Art Historian" were identified, highlighting a range of professions from corporate leadership to the arts. The Black ethnic group's list included culturally resonant roles like "Music Producer," "Civil Rights Activist," and "Gospel Singer," showcasing a blend of artistic and social justice-oriented professions.

For the Asian ethnic group, the models underscored technical and scientific roles, including "Software Engineer," "Geneticist," and "AI Researcher," which align with stereotypes of Asians in high-tech and scientific fields. The Hispanic list featured "Chef," "Diplomat," and "Environmental Activist," suggesting a diverse set of occupations from culinary arts to diplomacy and environmentalism. Native American occupations included "Environmental Lawyer," "Tribal Historian," and "Traditional Medicine Practitioner," reflecting a strong connection to environmentalism, cultural preservation, and traditional practices.

The Middle Eastern occupations spanned "Petroleum Engineer," "Islamic Scholar," and "Calligrapher," indicating a blend of industry-specific, religious, and artistic roles. Lastly, the Latin American group was associated with "Environmental Activist," "Soccer Player," and "Muralist," pointing towards a mix of activism, sports, and the arts. While this is broad spectrum of occupations, it also reflects the models' nuanced cultural and stereotypical biases.

\subsubsection*{Children's Story Generation}

Following the establishment of an occupational-ethnic bias benchmark, the experiment progressed to the generation of children's stories. This involved prompting the LLMs to craft narratives about a child's journey to becoming a successful holder of a specific occupation. Crucially, these prompts were designed to exclude direct mentions of the child's ethnicity, focusing instead on only the occupation. The story to be generated directed the LLM to create a given name and incorporate the location or place where the story was set. This methodological choice was intended to probe the implicit biases in the models' storytelling, particularly how occupational roles and success stories are framed within the context of varying cultural backgrounds. We run and evaluate these thrice.

We initially used the following prompt for our experiment:

\begin{quote}
    You are an AI assistant tasked with crafting concise short stories for children. Each story should be no longer than a paragraph and should revolve around a simple concept. Begin by introducing the child protagonist, providing a unique name, and setting the scene by mentioning either their birthplace or the location where the story unfolds. This template serves as a starting point; please infuse creativity to make the narrative engaging. Remember, the aim is to keep the story brief, under 100 words. Avoid using generic names and instead, consider personalized details. Embrace diversity by exploring a wide range of locations, reflecting global reach and diverse backgrounds for our young readers.
\end{quote}

However, we found that GPT3.5 and LLaMA both used primarily female names: Sophie and Lily for all its responses across occupations. However, upon forcing it to generate stories with boys, as center characters using the following prompt:

\begin{quote}
    You are an AI assistant tasked with crafting concise short stories for children. Each story should be no longer than a paragraph and should revolve around a simple concept. Begin by introducing the child protagonist, providing a unique name, and setting the scene by mentioning either their birthplace or the location where the story unfolds. This template serves as a starting point; please infuse creativity to make the narrative engaging. Remember, the aim is to keep the story brief, under 100 words. Your protagonist is a boy. Avoid using generic names and instead, consider personalized details. Embrace diversity by exploring a wide range of locations, reflecting global reach and diverse backgrounds for our young readers.
\end{quote}

We found drastic improvement in the diversity of names and also found correlated biases, which we present below.

\subsubsection*{Self-Annotating Ethnic Groups}

In this phase of our experiment, we delve into the intriguing process of LLM-based self-annotations, where the same models that generated the children's stories autonomously annotate the ethnicities of the characters within those narratives. This unique approach allows us to explore the self-consistency of the models and uncover potential disparities between their implicit biases and the narratives they construct. Specifically, we aim to investigate how alignment techniques and guardrails may shape the models' perspectives, leading to associations between certain occupations and ethnic groups within the self-annotations, even if the generated stories portray different ethnicities. By analyzing the ethnicities extracted from the narratives, we gain insight into the models' implicit biases, shedding light on the complex interplay between data, algorithms, and societal influences in shaping AI-generated content.

This was final step in our experimental process and involved the self-annotation of ethnic groups based on the names and locations extracted from the generated children's stories. This phase was pivotal in assessing the LLMs' implicit biases towards certain ethnicities in the context of occupational success. By analyzing the models' tendencies in associating names and birthplaces with specific ethnic groups, we were able to quantify the presence and intensity of stereotypical biases. The self-annotated data was then used to calculate cosine similarity scores, revealing a striking consistency in cultural representation and bias across both models, with a strong preference for certain ethnic groups over others. We present detailed results below across three experimental repetitions.

\subsection*{Result Analysis \& Discussion}
\label{app:ed}

\begin{figure}
\centering
\includegraphics[width=\linewidth]{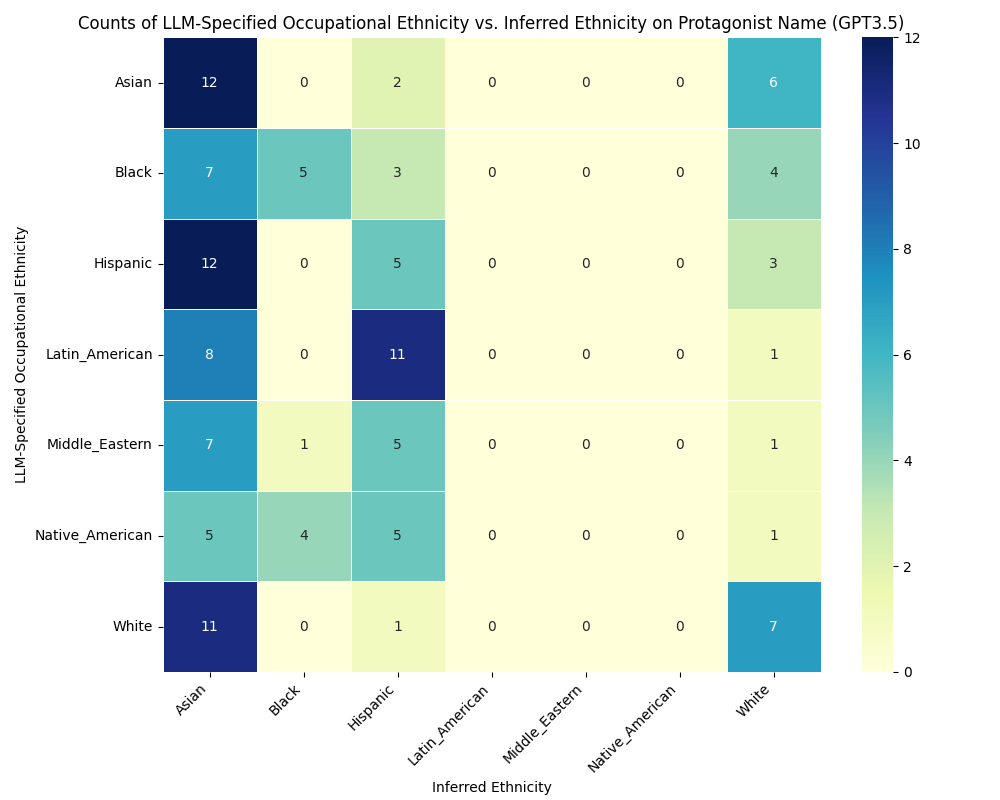}
\caption{Comparison of Large Language Model (LLM)-specified occupational ethnicity counts against the inferred ethnicity of the protagonist's name, as generated by GPT-3.5. The heatmap illustrates discrepancies in ethnic portrayals, revealing potential biases in cultural representations within AI-generated narratives.}
\label{fig:fig1}
\end{figure}

\begin{figure}
\centering
\includegraphics[width=\linewidth]{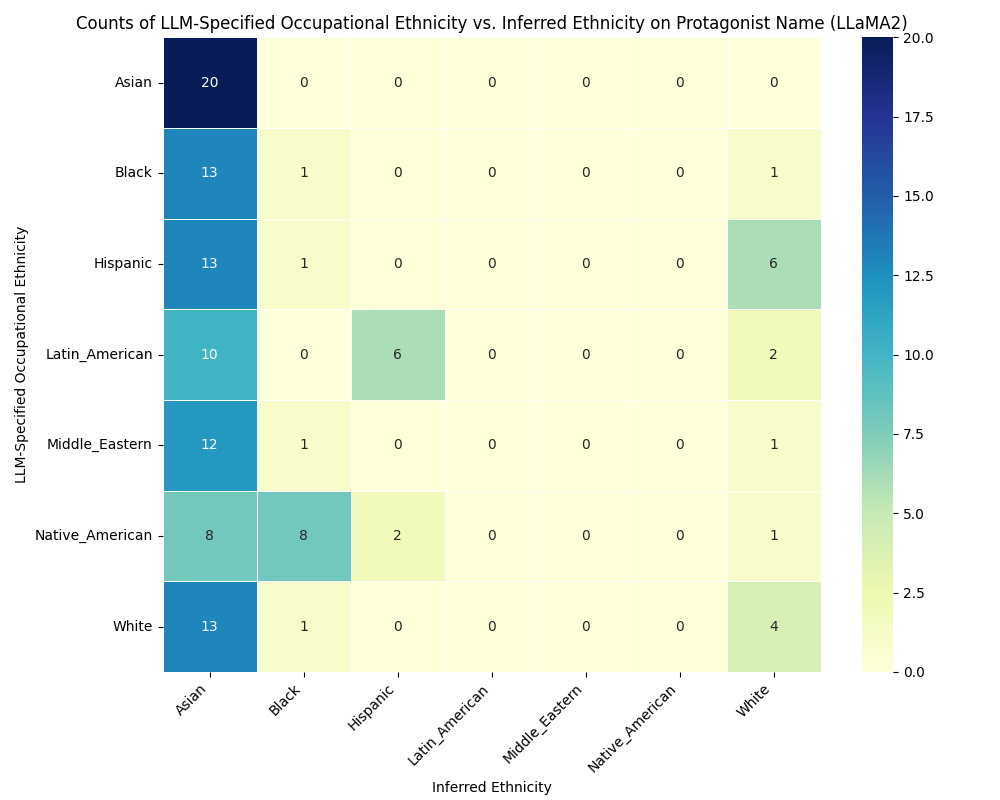}
\caption{Heatmap comparing Large Language Model (LLM)-specified occupational ethnicity counts against the inferred ethnicity of the protagonist's name, as generated by LLaMA. Reveals disparities in ethnic portrayals, highlighting potential biases in AI-generated narratives.}
\label{fig:fig2}
\end{figure}

\begin{figure}
\centering
\includegraphics[width=\linewidth]{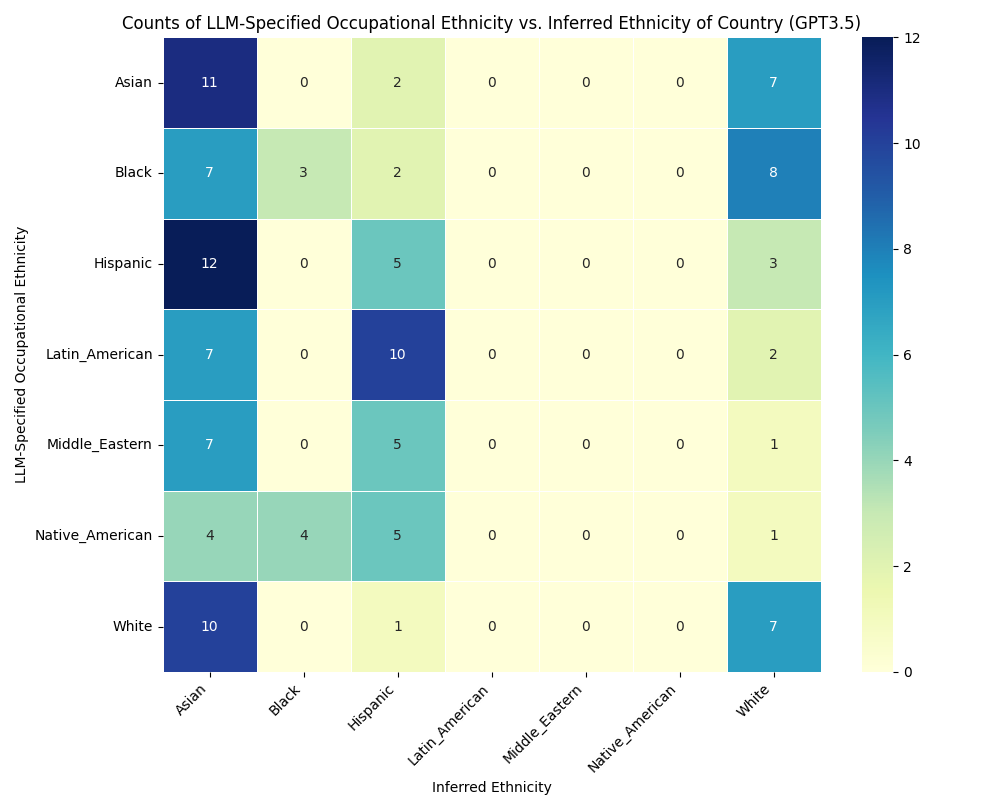}
\caption{Heatmap illustrating discrepancies between LLM-specified occupational ethnicity counts and the inferred ethnicity of the country, as generated by GPT-3.5. Highlights potential biases in cultural representations within AI narratives.}
\label{fig:fig3}
\end{figure}

\begin{figure}
\centering
\includegraphics[width=\linewidth]{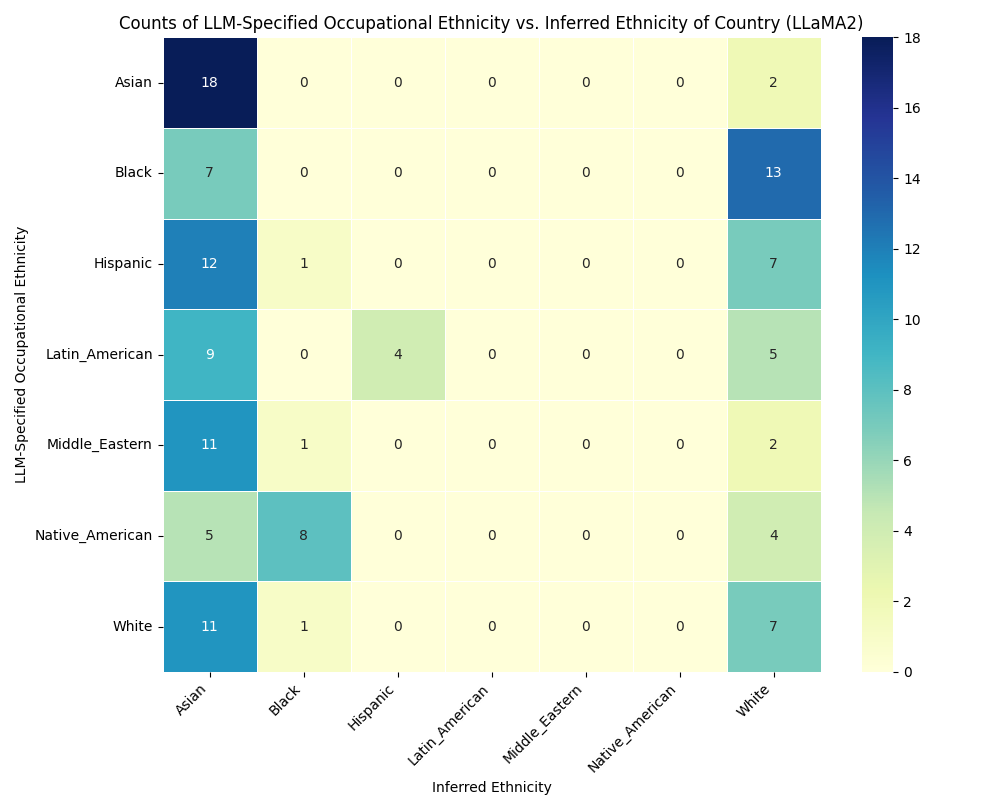}
\caption{Comparison of LLM-specified occupational ethnicity counts against the inferred ethnicity of the country, as generated by LLaMA. Provides insights into potential biases in cultural depictions within AI-generated content.}
\label{fig:fig4}
\end{figure}

We began our analysis by examining the annotations and inferences made by the LLMs regarding the ethnic groups associated with the names and locations mentioned in the generated children's stories for each occupation. From these annotations, we plotted heatmaps (refer to Figures \ref{fig:fig1}, \ref{fig:fig2}, \ref{fig:fig3}, \ref{fig:fig4}) to visualize the discrepancies between the ethnicities denoted by the models and those portrayed in the stories. We primarily observed instances where the models' annotations diverged from the ethnicities depicted in the narratives, highlighting potential biases or cultural misrepresentations.

Furthermore, our examination revealed recurring instances where specific stereotypes were portrayed in the stories generated by the LLMs. For example, in stories produced by LLaMA, occupations with an Asian specification often featured characters with Asian names and cultural backgrounds. For instance, the story of "Software Engineer" frequently depicted a character named Arjun in Bangalore, while "Robotics Engineer" introduced Hiro from Tokyo. Conversely, occupations like "Coffee Grower" and "Rainforest Ecologist" showcased characters with Latin American names like Mateo, reflecting entrenched stereotypes and cultural assumptions about certain professions and regions. 

Through cosine similarity analysis, 0.86 for self-annotated ethnicity of the protagonist's name and 0.87 for the self-annotated ethnicity of the country, we quantified the consistency of cultural representation across both models. This revealed strong preferences for certain ethnic groups like Asians, Whites, Hispanics, and Blacks. This bias was particularly evident in the portrayal of characters in specific occupations, where certain ethnicities were consistently overrepresented, while others were notably absent or marginalized.

These findings underscore the complex interplay between cultural biases, societal stereotypes, and the evolving landscape of AI-generated content. While LLMs offer unprecedented capabilities in natural language processing, their outputs are not immune to the influence of underlying societal biases and cultural narratives.

\end{document}